# A stakeholder-oriented multi-criteria optimization model
# for decentral multi-energy systems

*Using a nested multi-level decomposition approach*

Nils Körber, Paul Maximilian Röhrig, Andreas Ulbig

Institute for High Voltage Equipment and Grids, Digitalization and Energy Economics (IAEW), RWTH Aachen University, 52056 Aachen, Germany

**Contact:**
Nils Körber, M.Sc. | Tel.: +49 241 80-93051 | n.koerber@iaew.rwth-aachen.de | www.iaew.rwth-aachen.de

## ABSTRACT

The decarbonization of municipal and district energy systems requires economic and ecologic efficient transformation strategies in a wide spectrum of technical options. Especially under the consideration of multi-energy systems, which connect energy domains such as heat and electricity supply, expansion and operational planning of so-called decentral multi-energy systems (DMES) holds a multiplicity of complexities. This motivates the use of optimization problems, which reach their limitations with regard to computational feasibility in combination with the required level of detail. With an increased focus on DMES implementation, this problem is aggravated since, moving away from the traditional system perspective, a user-centred, market-integrated perspective is assumed. Besides technical concepts it requires the consideration of market regimes, e.g. self-consumption and the broader energy sharing, within a community. As the EU is increasingly pushing for the latter in terms of energy communities, this shows the dependency of such a perspective on the energy policy framework. Further it highlights the need for DMES optimization models which cover a microeconomic perspective under consideration of detailed technical options and energy regulation, in order to understand mutual technical and socio-economic and -ecologic interactions of energy policies.

In this context we present a stakeholder-oriented multi-criteria optimization model for DMES, which addresses technical aspects, as well as market and services coverage towards a real-world implementation. The current work bridges a gap between the required modelling level of detail and computational feasibility of DMES expansion and operation optimization. Model detail is achieved by the application of a hybrid combination of mathematical methods in a nested multi-level decomposition approach, including a Genetic Algorithm, Benders Decomposition and Lagrange Relaxation. This also allows for distributed computation on multi-node high performance computer clusters.

## 1. INTRODUCTION

### Decarbonization and Technical Solutions

Tackling the decarbonization of municipal and district energy systems – places where people live and work – is a key issue for achieving climate neutrality. Within the scope of municipal and district energy systems are the residential and commercial (CTS) domains, strongly connected to the building sector. Around 65% of the primary energy consumption worldwide can be attributed to those areas and domains, yet fossil fuels are mostly consumed. The basic end energy demands are first and foremost apportioned to room heating (and cooling), electricity-based applications, and warm-water supply. [1, 2] Towards a sustainable development of such energy systems, the substitution of fossil fuels, vice versa the integration of renewable energies, and the reduction of primary- and end-energy consumption are essential. In this matter, various measures need to be implemented, such as retrofitting the building stock by insulation reinforcement, the exchange or additional installation of electricity and heat supply technologies, as well as the installation of energy storage systems. [3, 4] While these measures are usually targeted towards an individual building transformation, there is increasing discussion on decentral multi-energy system concepts that offer cross-building and cross-sector efficiency on a larger, but local scale (see [5–8] ). A DMES concept is characterized by the optimal integration of different energy sectors (such as heating and electricity) via sector coupling. Therefore DMES comprise energy generation, distribution and demand in populated areas and their near proximity [9], as outlined in Figure 1.

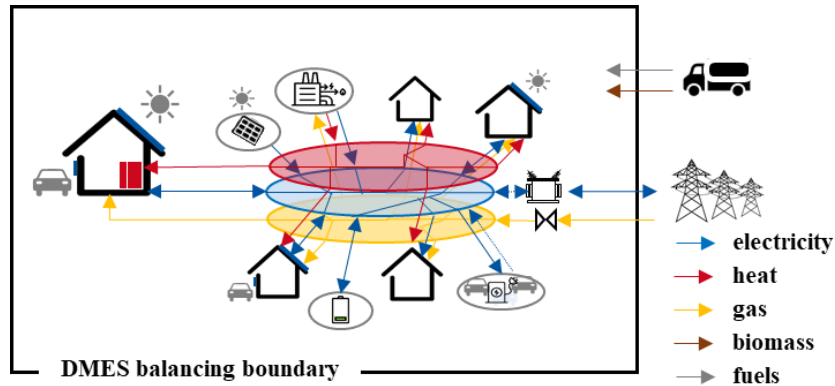

Figure 1. General Structure of a DMES

The balance boundary of the system borders the geographical area of such a system, which can consist of several dozen to a hundred buildings that form a district of a city or municipality. The highlighted cross energy carrier and infrastructure approach offers manifold synergies since it allows for a more rational use of local produced favorable renewable energy and efficient utilization of flexibilities in operation [5, 10, 11]. Additionally, restrictions for self-consumption on the building level, e.g. limitations in the available installation space or the available energy sources by smart energy distribution can be circumvented. Therefore, introduction of information and communication technologies (ICT) for automation of such systems in the sense of microgrids plays a major role.

### Consumer Centric Energy System Paradigm

The above described technical decarbonization approaches under the scope of DMES illustrate the inherent complexity and diversity of an energy system that will be highly decentralized in the future. In the contemporaneous context of decarbonization and decentralization, also democratization of the energy supply is more and more relevant. The concept of energy democracy is recognizing citizens' initiatives as counter currents to centralized energy decision-making [12] and underlines the need for a consumer centric energy system paradigm [13]. This is addressed by the European Commission in the Clean Energy Package: „„[…] reform the energy market to empower consumers and enable them to be more in control of their choices when it comes to energy. […] For citizens, it means better information, possibilities to become more active on the energy market and be more in control of their energy costs. […]," [14] With respect to DMES, the initiatives of renewable energy communities (RECs) and the linked energy sharing (ES) supported by European legislation (see [15, 16]) represent an important pillar of this new paradigm. In the EUs conception RECs are seen as the organization of local (non-utility) stakeholders that undertake joint activities – establishing collaborative energy ownership and supply – for the purpose of transforming the local energy supply towards a sustainable, climate friendly energy system. In this coherence, the promotion of RECs is expected to increase public acceptance, mobilize private capital and increase flexibility in the market [16].

### Motivation for stakeholder-oriented DMES optimization

As outlined previously, consumers are a major contributor to the local energy transition, which is an important factor in a holistic transformation strategy throughout Europe. Their investment activities in the design of a sustainable self-centred energy supply systems in order to become prosumers have major significance. This raises important questions for a) the stakeholders themselves and b) the regulators.

*a) Stakeholders:* From a stakeholders perspective decision making is complex. The beforehand-mentioned design philosophy of DMES offers a wide range of configuration options for the technical design and operation of the system. Besides, decisions may not only be driven by economic motivation but also by ecological nature. Generally the energy political framework in the form of, e.g. subventions, taxes, and bans, has a strong influence on the basis for these decisions. A major question is: What is the optimal techno-economic and ecologic design of a DMES under consideration of the energy policy framework?

*b)  Regulators:* In order to achieve the decarbonization targets, the energy policy framework formulated by the regulators are required to provide appropriate incentives. Thus, in order to assess the impact of legislation on consumers' long-term investment behavior, it is necessary to put oneself in the consumers' perspective. The major question is vice-versa: What is the influence of the energy policy framework on the optimal techno-economic and -ecologic design of DMES?

### *Literature review*

DMES optimization belongs to the superordinate class of Energy System Optimization Models (ESOMs), which can be classified by focus area/ spatial dimension and their contextual resolution [17, 18]. DMES in our understanding are ESOMs with small scale spatial dimension representing the focus area of cities and districts. The properties of their contextual dimension comprise: spatial resolution (consumer sharp, house sharp, block of houses), temporal resolution (milli seconds – years), time horizon (days – years), sectoral coverage (electricity, heat, cooling ,fuels, etc.), network coverage (electricity, gas and heat networks), technical coverage (renewable generation, storages, etc.), as well as market and services coverage (self-consumption, energy sharing, etc.) [19–21].

In this paper, we further distinguish models with a central planning perspective and models with a stakeholder oriented DMES optimization approach. The former models are widely represented in the DMES common body of literature [22]. The central planner approach focuses primarily on the technical aspects of DMES, as well as their economic and ecological dependencies. Yet market roles, consumer choices, and non-regulated and regulated infrastructures (district heating vs. electricity and gas networks) are conceived collectively. Exemplary for this are models like van Beuzekom [23], Murray [24], urbs [25, 26] and KomMod [27]. Stakeholder-oriented models, on the other hand, such as Falke [28], MANGO [29], RE$^3$ASON [30] and DER-CAM [31] reflect the decision-making of specific roles such as end consumers and housing associations, thus neglecting certain DMES optimization decisions while introducing market and services coverage. In this context, the roles and decision-making competencies are governed by the regulatory framework. Koirala [32] adapts the stakeholder oriented approach to energy communities, broadening the market and services coverage of such models. On the other hand the energy community approach in [32] is limited to certain technological options, focusing on the electricity supply.

### *Contribution of this paper*

Building on the common body of literature this paper presents a model which combines the strength of existing stakeholder-oriented models, to broaden the scope of DMES models in their contextual scope. Allowing for high spatial resolution, broad sectoral coverage, extensive technical coverage and market and services considerations, with a focus on the energy community approach. To cope with complexity in the respective optimization problem, a hybrid combination of mathematical methods is applied to avoid several simplifications, while still satisfying computational feasibility and holding a required model level of detail. The presented method uses a nested multi-level decomposition approach by combining evolution-based heuristics, dual dynamic programming, and cross-decomposition techniques – Genetic Algorithms (GA), Benders Decomposition (BD) and Lagrange Relaxation (LR)). This also allows for distributed computation on multi-node high-performance computer clusters.

The remainder of the paper is structured as follows: Section 2 introduces a general problem formulation in order to clarify the entire complexity of the approach in a comprehensible way. Based on this, section 3 presents the method and the mathematical problem formulation. At last, section 4 provides a summary and outlook for future work.

## 2.  GENERAL PROBLEM FORMULATION

In the following, the general problem formulation is presented. For this purpose, this formulation follows the structure of the mathematical modelling with respect to the objectives, the degrees of freedom and the constraints, in order to translate the real problem into a sufficient algebraic structure.

*Objectives*

In general ESOMs explore objectives in the triangle of energy economics between the conflicting interests of economics, ecology, and technology / security of supply. [19] While technology / security of supply are usually severe restrictions and basic requirements for a system design (see also subchapter *Restrictions*) in modern Europe, especially economic and ecologic concerns are conflicting areas. Those conflicting areas are addressed in our approach and represented by the annual total expenditures for energy investments and operations ($TOTEX^{Ann}$), as economic benchmark, and the climate impact measured in annual $CO_2$ equivalents based on the full life cycle of assets/ materials and energy carriers used ($CO2E^{Ann}$), as ecologic benchmark. The $TOTEX^{Ann}$ cover annualized capital costs ($CAPEX$) - expenses for construction and commissioning -, and annual operational costs for insurance, maintenance, commodities and electricity ($OPEX$). The stakeholder perspective includes all induced regulatory parameters, such as subventions, premiums, and taxes in cost calculation. $CO2E^{Ann}$ comprise relevant Green-House-Gas (GHG) Emissions from mining and processing of raw materials, production of components and equipment, and their composition, use of auxiliary energy during transport, combustion of commodities and recycling.

*Degrees of Freedom*

In terms of a cellular system within the overall energy system, local energy generation within a DMES is rather complementary to the facilities and infrastructures outside the balance boundary. However, the role of local energy supply is becoming increasingly relevant for the decarbonization goals. For this reason, all ubiquitous renewable energy sources (RES) will need to be exploited wherever possible. At the same time, the generation of energy close to consumption offers advantages in efficiency, since energy does not have to be transported and distributed over long distances. An overview of the local design options that are covered in our model for DMES are illustrated in Figure 2 and 3.

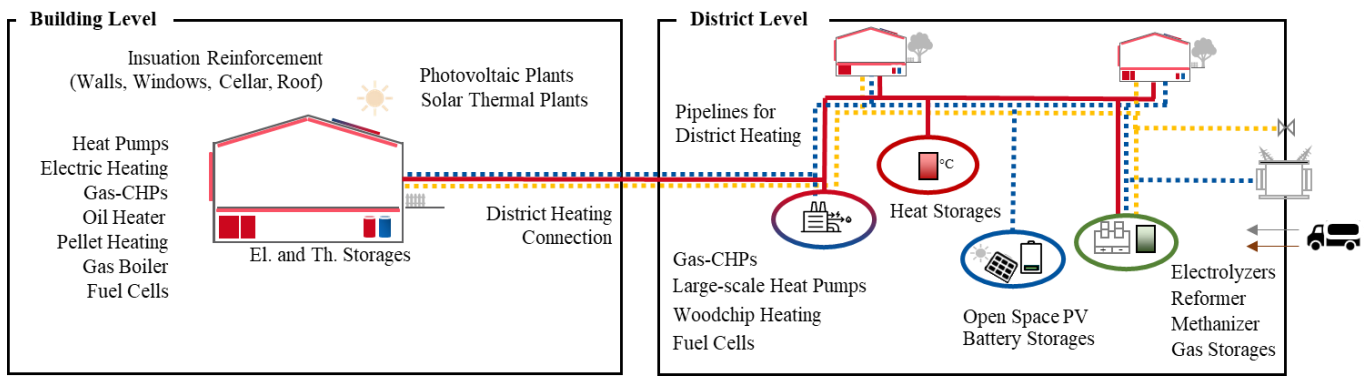

Figure 2.   Design options for DMES on building and district scale

Figure 2 emphasizes the nature of investment decisions which lay on individual building level and community district level. Figure 3 highlights' the cross-energy-carrier, sector-coupled approach of DMES, wherein energy conversion technologies ($ECT$) provide the gateways. Overall, a variety of ECT can be installed that allow the use of multiple energy resources. Further on energy storages ($ESS$) represent design options, which allow for efficient system operation. While gas and electricity grids in modern Europe are usually widely pre-existent, district heating networks ($DHN$) may also be new design options. The reinforcement, the expansion, or also deconstruction/rededication (for H₂) of the politically regulated grids for electricity and gas are not considered in the present model. In addition to the pipe layout and sizing for DHN the connection of a building to a DHN, the heat transfer in the form of heat exchangers ($HE$) is subject to degree of freedoms in the model. Besides the technologies and infrastructure, also insulation reinforcement ($IR$) of buildings decreasing the heat demands is a design option.

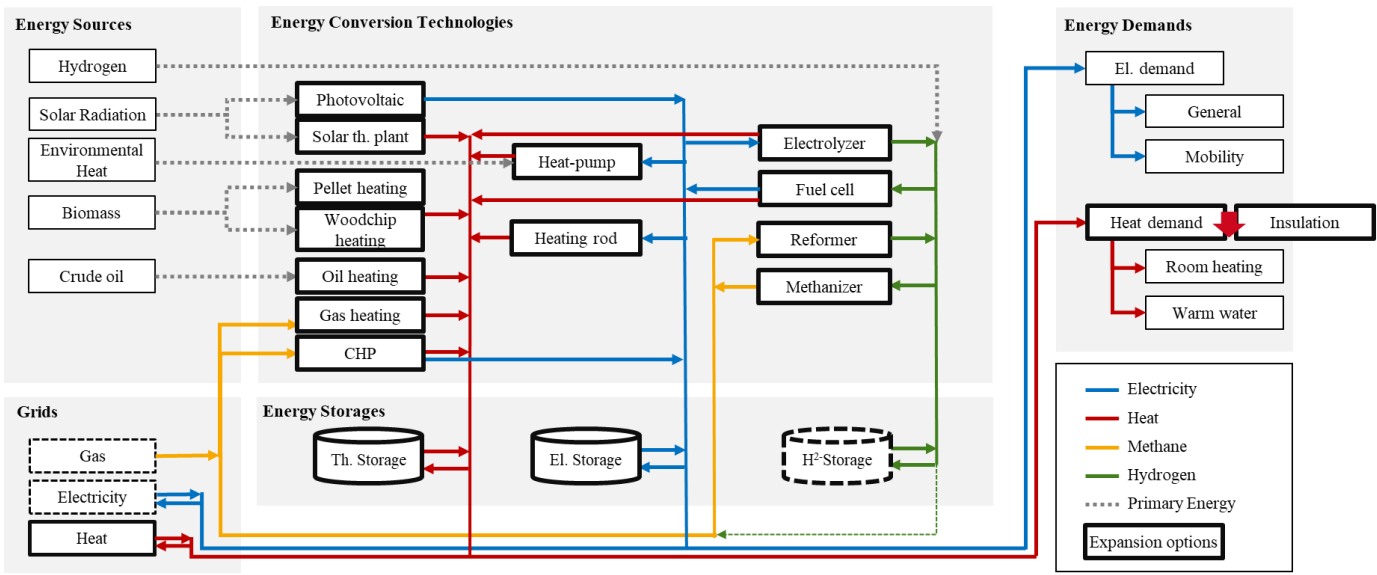

Figure 3. Design options for DMES to cover the basic energy demands.

A holistic energy-efficient design of DMES also covers operation aspects, as expansion and operation have direct dependencies, e.g. due to the conditional dimensioning of installed power of the ECTs or capacities of the ES. Operation in our terms covers technical operation, as well as market-oriented operation being the overlying layer from an end consumer/user-perspective. From a DMES point of view technical operation encloses a combined operation of assets to serve the heating and electricity demands. The freedom of degrees of ECT are energy in- and outputs, which are in direct relation due to the technical behaviour of the ECT. Additional operation freedoms for ES lies in the storage option, where fill level/state of charge can be optimized.

From a market perspective distributed generation requires the coordination of assets owned by multiple stakeholders. Instead of direct control, as in a technical operation, market signals incite coordinated behaviour. These market signals are manifold in modern energy systems and are prompted by various marketing and procurement options for energy and commodities. An essential element on decentral scale is on the one hand the classic procurement of energy via supply contracts with energy utilities. On the other hand, self-consumption poses a consumer centric (prosumer) opportunity. The counterpart to self-consumption is the direct marketing of self-generated energy. Between the classic energy procurement via contracts and the self-consumption/ direct marketing, energy sharing is a new way of exchanging renewable energy in the legal entity of an energy community. Energy sharing might be exercised within a peer-2-peer or centrally coordinated local energy market as described in [33–36]. We assume a perfect market with no information asymmetries; thus, the outcome of both structures yields the same results, which is also reflected in our modelling approach (See: Lagrange Relaxation vs. Closed Optimization). In this context, besides pure technical degrees of freedom, also the essential marketing options are covered.

### *Restrictions*

The constraints/ restrictions ensure that the DMES energy supply is designed in a practice-oriented manner. Therefore, this modelling aims to represent the in-depth limitations and dependencies in investment and operational planning. At the invest stage – also addressed as "Here-and-Now" stage, constraints are mainly related to the mutual exclusiveness of decisions (e.g. exclusivity of technologies). Also, maximum plant sizes due to the limited available space are considered. An overall key constraint is the coverage of energy demands. In direct relation stands the proper dimension of assets, as well as operation must be ensured. In this context, dependencies between renovation measures and the heat demand are also modelled. Further restriction on the operational level – also addressed as "Wait-and-See" stage, ensure that the technical capabilities of the plants are not violated (e.g. maximum power limits). Besides technical operation constraints, also the market layer is considered. Further constraints are imposed by the regulatory framework, e.g. the prohibition of certain technologies, or the limitation of the spatial expansion of an EC.

## 3. ALGORITHM AND MATHEMATICAL PROBLEM FORMULATION

### Decomposition of Decision-Making

The algorithm is based on multiple decomposition concepts, which divide the originally large overall problem into relatively smaller interdependent subproblems. This involves decomposing the problem in both the time and system domains within a nested superstructure. The chosen decomposition approach is illustrated in Figure 4.

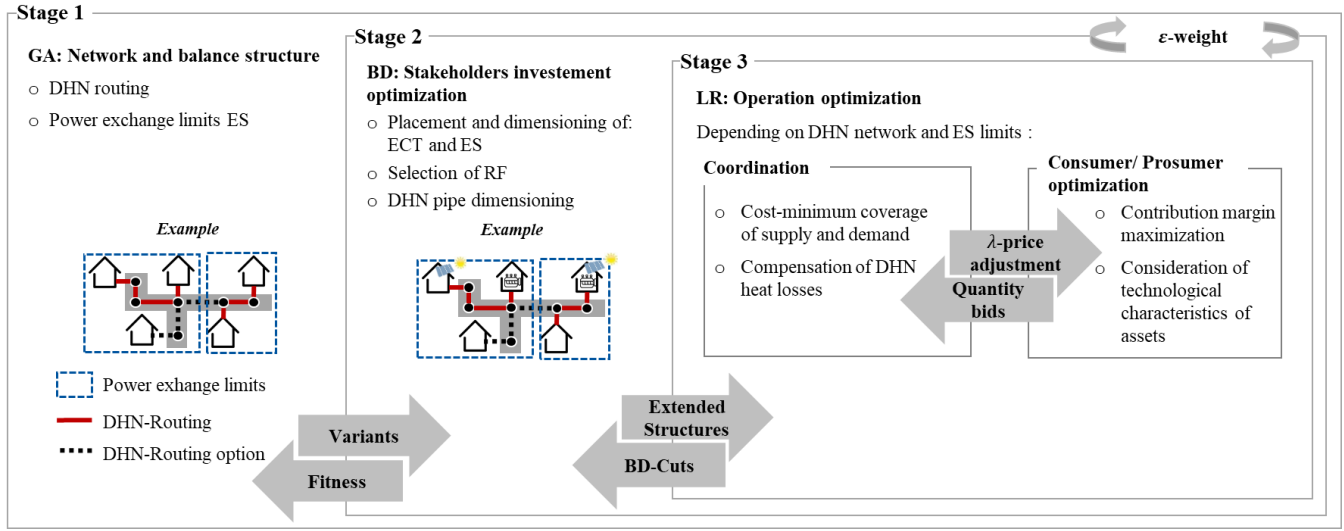

Figure 4.   Nested algorithm structure

At this point, we will only briefly discuss the decomposition of decision making in this approach. The features of the individual stages will be discussed in more detail later. At the first stage of the model, a GA is applied to optimize the local heating network routing and the limits of a local energy exchange (with limited geographic scope for expansion). As shown in the example, one individual of the GA represents a possible variant of these structural properties of a DMES. Within the fitness evaluation for the different variants in one generation of the GA a separate mixed-integer linear optimization problem (MILP) is constructed for each coherent sub-structure per individual of the GA. This optimization problem includes the expansion and operation of the remaining freedoms of degree. On the left-over expansion stage this involves the dimesioning of DHN pipes and all ECT, ES and refurbishment decisions. With Stage 2 (the BD) the remaining expansion decisions are dissociated from the operating decisions in a respective MILP. Under a BD iteration, the "here-and-now" decision thus predetermines an extended structure of the DMES for operation optimization. This stage 3 can again be decomposed into independent subproblems by LR. This is only the case if several buildings are connected to each other via previously predefined structures. The LR allows for a relaxation of the connected load coverage constraint, thus a separation of the buildings and their operational optimizations. Overall optimality on the LR stage is guaranteed by an iterative coordination with quantitiy bids and a following market-clearing via converging shadow-prices of the con- and prosumer. Optimality at the BD level is likewise achieved by the integration of a mathemtical estimator in the "here-and-now" decision, which converges using so-called BD cuts per iteration. The genetic algorithm as the top layer does not necessarily pursue the goal of finding the optimum of the decisions anchored at its stage. More importantly, it serves to determine a practical preselection of possible network and balance structures. The nested structure of the decomposition is implemented programming-wise according to the flow chart shown in Figure 5. A description of it and its mathematical components is given in the following subsections.

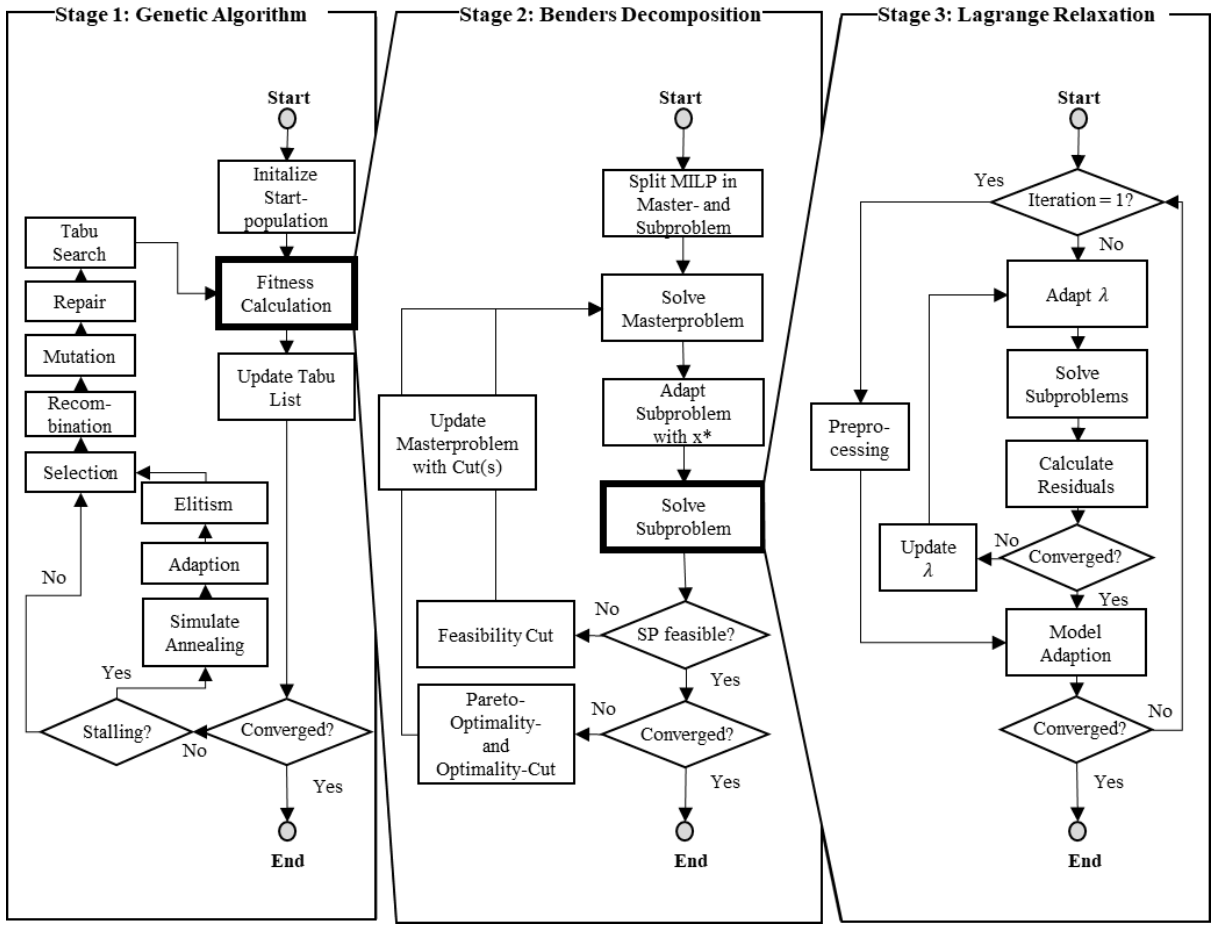

Figure 5.   Flow Chart of the Algorithm

### Stage 1: Genetic Algorithm

A GA belongs to the class of stochastic, metaheuristic optimization methods and is based on the theory of evolution. Our GA approach is inspired by the NSGA (nondominated sorting genetic algorithm) -II and -III [37] , which is applied to multicriteria optimization problems. Also, the work of T. Falke [38] served as a foundation.

#### 1)   Start population

The GA process starts with a start population as first generation, representing several design options for a DMES as individuals of the population. The genetic encoding of the design options is outlined in Figure 6.

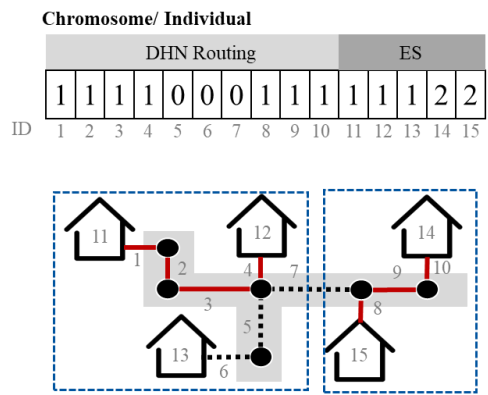

Figure 6.   Genetic encoding of design options covered in the GA

A chromosome as equivalent to an individual in the population consists of two separate sections. 1.) The first section encodes the routing of the DHN. The required route data is available to the model in georeferenced format. A gene in this section thus carries the information as to whether a specific pipeline is built or not built - a binary decision. The chosen encoding has the disadvantage that genetic defects can occur, which happens, for example, in the case that interrupted DHN are created by the random mutation. This is prevented by a repair function, which could also be implemented equivalently through a penalty cost function as part of the fitness evaluation. 2.) The second section encodes the spatial expansion of energy sharing. Here, a gene associates a group with each building in the area of the DMES – a discrete decision. Building are also available to the model in georeferenced format. Similarly, to the first section, genetic defects can occur within random mutation. This can be the case when groups with more than one associated building do not have a spatial context but is similarly prevented by the repair function. Even though there might be a spatial context, a broad spatial expansion of the balancing structure might also be the reason for a genetic defect. This is measured by an adjustable threshold that restricts the spatial diameter of the groups, thus representing restrictions by the energy political framework.

The start population in the present case consists of an adjustable size of random chromosomes and several predefined chromosomes. The predefined chromosomes being a combination of the extreme structures (no DHN, full DHN/ no ES, largest ES distribution in groups).

### 2) Fitness evaluation

Within the GA first a new generation is assessed for its fitness. The fitness evaluation consists of two parts. Both employ the BD, the first being a pre-calculation for the maximum mass flow per DHN pipe section, in order to find a suitable dimensioning of pipes and to derive respective grid losses. In this pre-optimization, only the heat side of the system is considered and the determination of the maximum heat flow $\dot{Q}_l^{max}$ is understood as a transport problem based on the optimized heat supply task.

The maximum occurring mass flows $\dot{m}_l^{max}$ per pipe section $l$ are calculated according to Formula 1, with the specific heat capacity of the heat transfer fluid $c_p$, as well as the spread between flow $T_{FL}$ and return $T_{RF}$ temperature. The supply and return temperatures are assumed to be constant with a temperature spread of 25K. The flow temperature is set 20 K above the maximum flow temperature of all consumers connected to the network. [38]

$$\dot{m}_l^{max} = \frac{\dot{Q}_l^{max}}{c_p \cdot (T_{FL} - T_{RF})}$$

Formula 1.

The maximum mass flows are finally used for the dimensioning of the pipelines with nominal widths of DN 20-200 and corresponding mass flows of 0.5 -193 t/h. The loss per pipe is determined in the following based on Formula 2.

$$\dot{Q}_{l,t}^{loss} = \left(\frac{T_{FL} - T_{RF}}{2} - T_t^{Earth}\right) \cdot U \cdot A_L$$

Formula 2.

Thermal power loss results from difference between the mean temperature in the DHN and the ground temperature $T_t^{Earth}$, as well as from the heat transfer coefficient $U$ and the outer surface of the pipes $A_L$.

In the course of the second optimization, the results of the pre-optimization are also included. Results of the fitness assessment are both annuity costs and the annual $CO_2$-equivalent emissions. The fitness finally allows the multicriteria evaluation of individuals, which is done in the context of selection.

### 3) Selection

Selection is a genetic operation that serves to select individuals that are mutated and recombined further during the procedure. For problems with multi-criteria fitness, the literature recommends using Pareto-efficiency as a selection criterion. According to [37], the process of selection is done in two steps: non-dominance sort and tournament selection. In the first step the fitness value of the individuals is checked for dominance. Based on these results, the individuals are ranked in ascending order. The individual with the best fitness is in rank one, the second-best fitness in rank two, and so on. In addition, the crowding distance (CD) is determined between individuals of the same rank. The underlying calculation is shown in Formula 3 and Figure 7:

$$CD_p = \sum_{n=1}^{2} \frac{\left(F_n(p+1) - F_n(p)\right) * \left(F_n(p) - F_n(p-1)\right)}{F_{n,max} - F_{n,min}}$$

$$\forall\, p \in P$$

Formula 3.

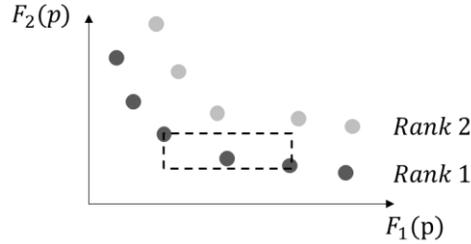

Figure 7.   Crowding Distance

The CD is the sum of the normalised distances between the fitness values $F_i(p)$ of the neighbouring solutions. Where $i$ in our cases represent the two fitness dimensions, economy and ecology, and $p$ is one individual in the whole population $P$. Overall CD is a measure of diversity within the population, by determining the proximity of the solutions fitness to all adjacent solutions fitness. Based on this, a competition selection takes place in the second step. For this purpose, two random individuals are selected from the population and compared according to their rank and CD value. The rank serves as the first basis for decision-making. The individual with the lower rank is selected. In case two candidates have the same rank, the individual with the higher CD is chosen. The reason for this is that mainly non-dominated individuals with a large distance are obtained. In this way, a broad consideration of the solution space is made. This selection procedure is intended to prevent the algorithm from converging to a local optimum. [37]

*4)   Recombination and Mutation*

To create diversity in the options to be assessed recombination and mutation is applied based on a fixed number of selected individuals from the previous generation. Recombination happens by first disassembling the chromosomes into their two essential sections and second recombining these sections. Mutation happens completely random, making the aforementioned repair algorithm necessary.

*5)   Tabu-Search*

The tabu-search was implemented according to the methodology of [39]. The process resembles human immune systems, where the taboo list is the system's antibodies. This involves creating a list that contains the candidates that are not allowed to be used in the further search for a solution. The solutions that have already been examined are added to this list for each iteration in our case. As a result, solutions from previous generations are not considered again. After the creation of a new individual, it is checked whether it is included in the list or not. If the individual is in the tabu list, the individual is discarded and a new one is created using the genetic operators explained above. The tabu search procedure thus avoids unnecessary and time-consuming recalculations of already examined individuals. These individuals would not improve the population further, as their solution has already been investigated. The procedure allows a concentration on new solutions and enables an accelerated solution search with higher diversity.

*6)   Simulated Annealing*

Simulated annealing describes a local search procedure that accepts a bad individual in the near vicinity of the best individual on probability-based acceptance in order to leave the local optimum of the algorithm. This can happen when no improvement in the solution has been achieved over generations – the GA is stalling or stagnating. The basic idea of simulated annealing is the higher probability of a transition from higher to lower minima in a minimisation problem than in the opposite direction [40]. Depending on a certain probability $(PB_{SA})$ a new candidate is introduced in the new generation instead of the overall best candidate. This probability depends not only on the fitness difference between the current best $F_{CB}$ and the chosen new candidate $F_{NC}$, but also on a temperature parameter $T_0$ and the "freeze" time of the algorithm, as in Formula 4.

$$PB_{SA} = l^{-\frac{F_{CB}-F_{NC}}{T_0}}$$                                                                                          *Formula 4.*

The parameter $l$ describes the current time step and $T_0$ a constant initial temperature. $T_0 = |F_{min}|$ is the minimum objective function value of the initial population. The $PB_{SA}$ is later compared against a randomly generated number between zero and one. In the case that a candidate has a higher probability, the individual is adopted into the population.

*7)  Adaptive Mechanisms*

Adaptive mechanisms (AM) refer to the adjustment of recombination and mutation probabilities based on the stalling degree [41]. The stalling degree indicates the time interval elapsed since the stagnation of the solution space. Similar to the simulated annealing, the search strategies of the algorithm are changed. In this mechanism, the best solution of each generation is measured. If no better solution is produced in a generation $G$, the generation $G_{frozen}$ in which the last best solution was found is fixed. The Formulae 3 and 4 describe how the recombination probability $PB_{RC}$ and the mutation probability $PB_M$ are adapted in this dependency, also illustrated in Figure 9:

$$PB_{RC} = PB_{RC0} + \frac{G - G_{frozen}}{G} * (\alpha - PB_{RC0})$$                                                            *Formula 5.*

$$PB_{M0} = P_{M0} + \frac{G - G_{frozen}}{G} * (\beta - PB_{M0})$$                                                               *Formula 6.*

The constant $G_{frozen}$ is a positive integer, $PB_{RC0}$ and $PB_{M0}$ are the initial probabilities of recombination and mutation. $\alpha$ and $\beta$ are control parameters of the AM. Typically, $\alpha$=0.9 and $\beta$=0.2 are chosen [42]. If stagnation is overcome due to a better solution found in the next generation, N is reset to zero and the probabilities are set to their initial values.

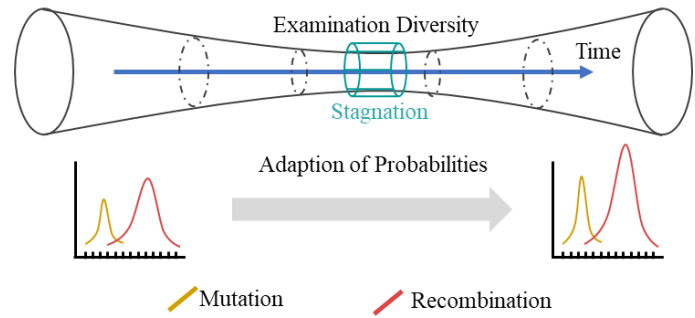

Figure 8.   Illustration of Adaptive Mechanisms

*8)  Elitism*

Elitism is based on Charles Darwin's evolutionary principle "Survival of the Fittest". The idea of using elitism founded on the idea of improving solutions as best as possible. Preference is given to individuals with better fitness so that their traits are passed on to their successors. There are three forms of elitism (see figure 9), which can be applied either or. Option (1) is always the adoption of the best candidate solution to guarantee an upper bound for the worst best solution per generation. Option (2) is a special case of Option (1), in which each generation always stores the best candidate so far. However, the candidate is only transferred to the population should stagnation occur. This limits the potential solution space to a lesser extent than in case one. However, it is possible to leave local optima again by means of the best solution. Option (3) includes the possibility of determining a new population only on the basis of the current best individuals, the so-called Pareto front, the first rank of fitness.

For this purpose, it is necessary to select only those parents from the population that lie within the first rank. These candidates are then combined and mutated using genetic operators. Elitism has the advantage that the overall fitness of a population does not deteriorate over the generation. In case three, the exclusive entrainment of the elite can lead to convergence in a local optimum. This must be prevented by other methods such as SA.

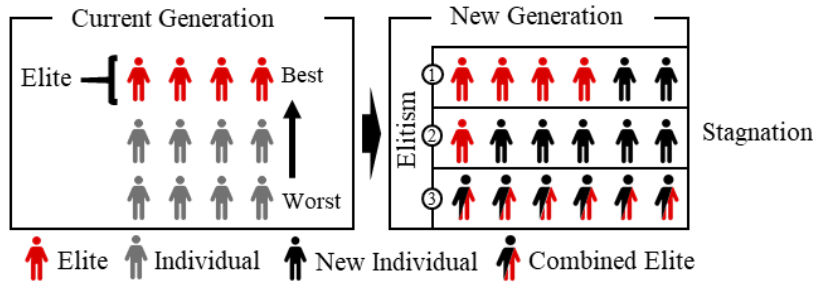

Figure 9.   Forms of elitism

### Stage 2: Benders Decomposition

Benders-Decomposition Technique (BD*), proposed by J. Benders in 1962 [43], is a procedure to solve large-scale MILPs or stochastic optimization problems with complicating variables [44]. For problems with a certain structure it promises the benefit of a shortage in computation time in comparison to solving a closed optimization formulation via conventional branch-and-cut or branch-and-bound methods. It is even possible to solve an unsolvable closed optimization problem with this cutting-plane technique [45]. BD has been applied and adjusted to a broad range of optimization problems, as described in [46]. The general concept of this approach relies on splitting the problem in two instances a master- and subproblems (MP, SP(s)) and fixing the interconnecting variables in the SPs by using their values from the master problems solution. To improve the overall solution reflected in the MP by the means of a better approximation of the SPs solutions, Benders cuts on the basis of dual variables of the obtained solutions in the SPs are introduced in the MP in every iteration of the Benders algorithm. [47] We apply the BD in the form of a two-step sequence, as a decomposition in the time domain. The decisions on investing are seen as "here-and-now" decisions and the operational optimization is seen as a "wait-and-see" decisions.

Initially, the problem is composed in a closed formulation, and later it is fragmented. This is due to the dynamic assignment of the problem structure from the superimposed GA. However, one advantage of this approach is that individual clusters in the DMES that result from the overlapping of DHN and ES groups can be constructed and solved separately. In some cases, individual buildings can be optimized completely separately from the rest of the system, if no interconnections exist. For these relatively frequently occurring repetitive cases, a database has been set up that contains solutions that have already been calculated. These can be accessed so that a second calculation of the same problem is not necessary.

#### 1)   MILP Model

Formula 7 represents the general objective function for the DMES optimization. Minimizing the total costs of energy supply $a^{TOTEX}$ and minimizing the resulting greenhouse gas emissions $e^{TOTAL}$ are two equally important but conflicting target criteria.

$$\min z = (1 - \varepsilon)(a^{TOTEX}) + \varepsilon(e^{TOTAL})$$

*Formula 7.*

The present approach to multi-criteria optimization uses the eta weight method, both targets being the weighted sum of the objective function with the weighting factor $\varepsilon$ in the continuous range of [0; 1]. The economic target comprises the annuities of the investment costs $a_l^{CAPEX}$ for the DHN pipes $l \, \epsilon \, L$, as well as the annuities of the investment costs $a_{b,i}^{CAPEX}$ per building b $\epsilon$ B and the available expansion measures $i \, \epsilon \, \{ECT, ES, IR, HE\}$, the energy conversion technologies $ECT$ , the energy storages $ESS$ , the insulation reinforcement $IR$ and the possible heat grid connection in the form of the heat exchangers $HE$ . In addition, the annuities of the operating costs $a_{b,j}^{OPEX}$ per building and respective technology $j \, \epsilon \, \{ECT, ESS, HE\}$ are part of the economic objective function, as shown in Formula 6.

$$a^{TOTEX} = \sum_{l \,\epsilon\, L} a_l^{CAPEX} + \sum_{b \,\epsilon\, B} \sum_{i \,\epsilon\, \{ECT,ESS,IR,HE\}} a_{b,i}^{CAPEX} + \sum_{b \,\epsilon\, B} \sum_{j \,\epsilon\, \{ECT,ESS,HE\}} a_{b,j}^{OPEX}$$    *Formula 8.*

The annuities of CAPEX for DHN pipes are calculated based on Formula 7, having fixed binary decisions $x_l^{fixed} \in [0,1]$, predetermined in the GA stage, and also fixed discrete decision variables for the dimensioning sizes $x_l^{dim} \in \mathbb{Z}$. Respective costs are length-dependent underground construction costs $c_l^{CAPEX,fix}$ and production/material costs dependent on lengths and pipe thicknesses $c_l^{CAPEX,var}$. $PVF_l$ represents the Present Value Factor for an assumed asset lifetime which is set at 40 years.

$$a_l^{CAPEX} = \frac{x_l^{fixed} \cdot c_l^{CAPEX,fix} + x_l^{dim} \cdot c_l^{CAPEX,var}}{PVF_l}$$    *Formula 9.*

The annuities of CAPEX for all other technologies and IR are calculated on the basis of Formula 8. For IR, the only cost occurred is in the decision itself $x_{b,i}^{bin} \in [0,1]$. Possible renovations are - wall, roof, windows, basement, and their combinations. Thus, there can be up to 16 different insulation options for a given insulation standard. The technological options for ECT and ESS are covered in Table 1. For both IRs and technologies, fixed CAPEX $c_{b,i}^{CAPEX,fix}$ are composed of fixed installation/construction and commissioning costs. This cost component also includes other costs for insurance, dismantling and disposal. Modelled subsidies, such as lump-sum payments by the government for the installation of renewable technologies, reduce costs. Variable cost items and proportional subsidies in variable CAPEX $c_{b,j}^{CAPEX,var}$ affects the dimensioning decision $x_{b,j}^{dim} \in \mathbb{Z}$.

$$a_{b,i}^{CAPEX} = \frac{x_{b,i}^{bin} \cdot c_{b,i}^{CAPEX,fix} + x_{b,j}^{dim} \cdot c_{b,j}^{CAPEX,var}}{PVF_i}$$    *Formula 10.*

Other relevant cash flows relate to demand-related (costs for fuel and electricity) and operations-related costs (costs for maintenance and repair). These variable costs $c_{b,j,t}^{OPEX}$ are recognized as part of the annuities of OPEX being related to operational decisions $y_{b,j,t} \in \mathbb{R}$, over a representative time horizon of a year $t \in T$ as shown in Formula 11.

$$a_{b,j}^{OPEX} = \sum_{t \in T} \frac{y_{b,j,t} \cdot c_{b,j,t}^{OPEX}}{PVF_j}$$    *Formula 11.*

The economic objective function $e^{TOTAL}$ is composed of annual $CO_2$-equivalent emission shares that are considered over the life cycle of the materials and energy sources used ("from cradle to gate"), as shown in Formula 12. On the one hand, these are emissions $e_{b,j}^{Operation}$ that can be attributed to the combustion of fuels in operation. On the other hand, these are emissions $e_{b,i}^{Other}$ that can be allocated to raw material extraction, processing, production, transport and disposal.

$$e^{TOTAL} = \sum_{b \,\epsilon\, B} \sum_{j \,\epsilon\, \{ECT,ESS,HE\}} e_{b,j}^{Operation} + \sum_{b \,\epsilon\, B} \sum_{i \,\epsilon\, \{ECT,ESS,IR,HE\}} e_{b,i}^{Other}$$    *Formula 12.*

Similar to the economic objective function, these annual emissions in operation $e_{b,j}^{Operation}$ are accounted by a factor for the greenhouse warming potential $gwp_{j,t}^{Operation}$ of fuel combustion (see Formula 13). Corresponding factors are also applied to the fixed and variable decisions related to the investment decisions (see Formula 14).

$$e_{b,j}^{Operation} = \sum_{t \in T} y_{b,j,t} \cdot gwp_{j,t}^{Operation}$$    *Formula 13.*

$$e_{b,i}^{fix} = x_{b,i}^{bin} \cdot gwp_i^{Other,fix} + x_{b,j}^{dim} \cdot gwp_j^{Other,var}$$    *Formula 14.*

Table 1: DMES system technologies and abbreviations

| Electricity | Heat | |
|---|---|---|
| Photovoltaic plants (*pv*) | Solarthermal plants (*sol*) | Brine water heat pumps (*bwp*) |
| Lead-acid batteries (*lac*) | Gas condensing boilers (*gcb*) | Air water heat pumps (*awp*) |
| Lithium-ion batteries (*lio*) | Heating rods (*elh*) | Pellet heatings (*peh*) |
| Combined-Heat-and-Power (*chp*) | | Woodchip heatings (*wch*) |
| Fuel cells (*fcl*) | | Buffer tanks (*ths*) |

| Other technologies | | |
|---|---|---|
| Electrolyzer (*elz*) | Methanizer (*mzr*) | Reformer (*rfr*) |
| Hydrogen storages (*h2s*) | Heat exchanger (*he*) | Electric grid connection (*eg*) |

The constraints for optimization involve a variety of technical and non-technological components. On investment stage essential restrictions are:

- The exclusivity of insulation reinforcement measures, as shown in Formula 15. The same applies to the installation of heating technologies (*elh, gcb, chp, elz, fcl, bwp, awp, peh, wch, he*) and battery storage systems (*lac, lio*).
- The availability of roof area $A_b^{Roof}$ of a specific building *b*, which limits the space for solar technologies $st \in ST$ (*pv, sol*), according to formula 16.
- Minimum and maximum nominal power $P_j^{min}$, $P_j^{max}$ of the ECT/ capacity of the ESS, due to existing sizes available on the market or the limitation in space of a specific building, as depicted in Formula 17. (Note: Several ECT are modelled in different sizes to account for non-linear piece-wise linear cost curves)

$$\sum_{ir \in IR} x_{b,ir}^{bin} = 1 \qquad\qquad \text{Formula 15.}$$

$$\sum_{st \in ST} x_{b,st}^{dim} \leq A_b^{Roof} \qquad\qquad \text{Formula 16.}$$

$$x_{b,j}^{bin} \cdot P_j^{min} \leq x_{b,j}^{dim} \leq x_{b,j}^{bin} \cdot P_j^{max} \qquad\qquad \text{Formula 17.}$$

In addition to restrictions on the level of the investment, secondary conditions exist to ensure the coupling between investment and operation decisions. Essential is the coupling of the dimensioning of assets $x_{b,j}^{dim}$ with the operational freedoms of degree $y_{b,j,t}$ (see Formula 18). The same applies to the dimensioning of DHN pipes and the respective pipe utilization.

$$y_{b,j,t} \leq x_{b,j}^{dim} \qquad\qquad \text{Formula 18.}$$

IR also has an impact on operation. Specifically, they lead to a reduction in thermal energy demand and thus have a direct influence on the thermal load coverage constraint of a building. Each *ir* has a time series $TL_{b,ir,t}$ which represents the specific heat demand for room-heating and warm-water over the period of one year in ¼ hour resolution. These are used as the basis for load coverage according to Formula 19, representing the restriction for a time step $t \in T$ over the whole horizon $T$, with the use of the Big-M method.

$$-Inf + TS_{b,ir,t}^{Thermal} \leq \sum_{ht \in HT} y_{b,ht,t}^{toTL} + x_{b,ir}^{bin} \cdot Inf \qquad\qquad \text{Formula 19.}$$

A similar constraint for the load coverage of the electrical demand exists purely in terms of operation, as shown in Formula 20. In contrast, however, only one time series $EL_{b,ir,t}$ is used here, since there is no link to the investment decision.

$$TS_{b,t}^{Electric} \leq \sum_{et \in ET} y_{b,et,t}^{toEL}$$

<div align="right">Formula 20.</div>

For the technical and non-technical modelling of the operation, separate sub models exist for each technology, which are represented by an input and output structure. The linking of input (*from*) and output (*to*) is done via technical models that represent the essential technical properties of energy conversion (efficiency guard, coefficient of performance) and storage (time- and SOC-dependent losses). In addition, there are sources that represent interfaces to the DMES superimposed system, e.g. the electric grid connection (*EG*). The modelling further includes slack variables with high penalty costs as part of the technology models that ensure solvability of the operational problem at all times. Altogether, the system is interlinked by constraints to the extent that energy flows can be resolved granularly, as shown in the upper section of Figure 10. Besides technical aspects, the operation model also includes a broader contextual scope of marketing operation. The auxiliary constraints for energy purchase and sale are adjacent to the building boundary of a building entity as shown in Figure 10. On the electrical side, the opportunistic sales and procurement options already mentioned in section 2 are considered. Depending on the regulatory design, the direct marketing of energy $y_{EG,t}^{toDM}$ pays different remuneration depending on the energy source and the size of the feeder, which is represented by a further differentiation. Electricity purchase from the energy supplier $y_{EG,t}^{fromEP}$ includes cost components that are also depending on the regulatory framework, including e.g. costs for grid usage and energy taxes. In the model, these can be charged both as a performance-based (per kW) price and as a unit-based (per kWh) price, thus allowing for an analysis of tariff systems. A special feature of the illustrated market operation is the energy sharing. In the context of modelling, the energy sharing can be understood as a set of decision variables (without directly assignable costs) and constraints that link the buildings to each other in the electric load coverage. This is implemented by means of a perfect market from which the required energy quantities $y_{EG,t}^{fromES}$ can be purchased or residual electricity quantities $y_{EG,t}^{toES}$ can be sold. The possibility of ES for a building entity is limited by the overlying GA. The ES is implemented via a balancing constraint, as shown in Formula 21.

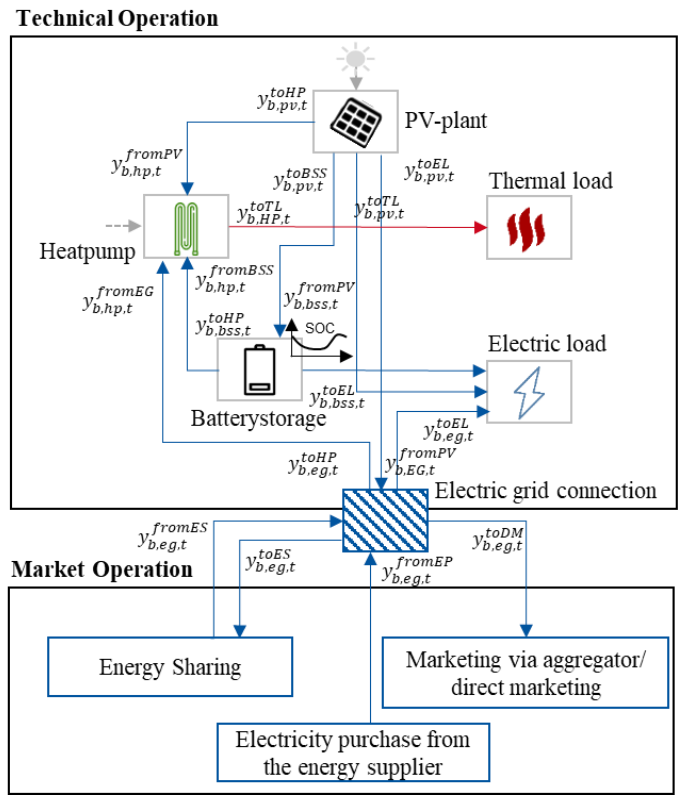

Figure 10. Illustration of the technical and market operational representation by the constraints

$$\sum_{b \in B} \left( y_{b,EG,t}^{toES} - y_{b,EG,t}^{fromES} \right) = 0 \qquad \text{Formula 21.}$$

$$\sum_{b \in B} \left( y_{b,HE,t}^{toDHN} - y_{b,HE,t}^{fromDHN} \right) - \sum_{l \in L} \dot{Q}_{l,t}^{loss} = 0 \qquad \text{Formula 22.}$$

In an equal form we consider the technical operation of district heating systems according to Formula 22, where in real world the grid dimensioning itself limits the transmittable energy through the lines. This is neglected at this point, since the line dimensioning and the losses of the network are subsequently/ or have been predetermined within the GA fitness evaluation.

### 2) Blockangular structure and decomposition of the MILP

The above described MILP is built within the algorithm as block-angular structure, as shown in Figures 11 and 12. The block-angular structure of a building as shown in Figure 11 comprises the Formulae relevant for the entity (see among others, Formulae 7 to 20). The relative larger structure for a coupling of buildings via DHN and ES is represented in Figure 12. In this respect, ndividual decision-making instances (investment and operation) of the building entities are separated from each other and reassembled in their components for the group. Coupling of the building entities in operation is done by Formulas 21 and 22. The structure of both block-angular structures is organized in such a way that a decomposition in the sense of Benders decomposition is possible. We use the notation according to formula 23, which represents the closed structure of the problem.

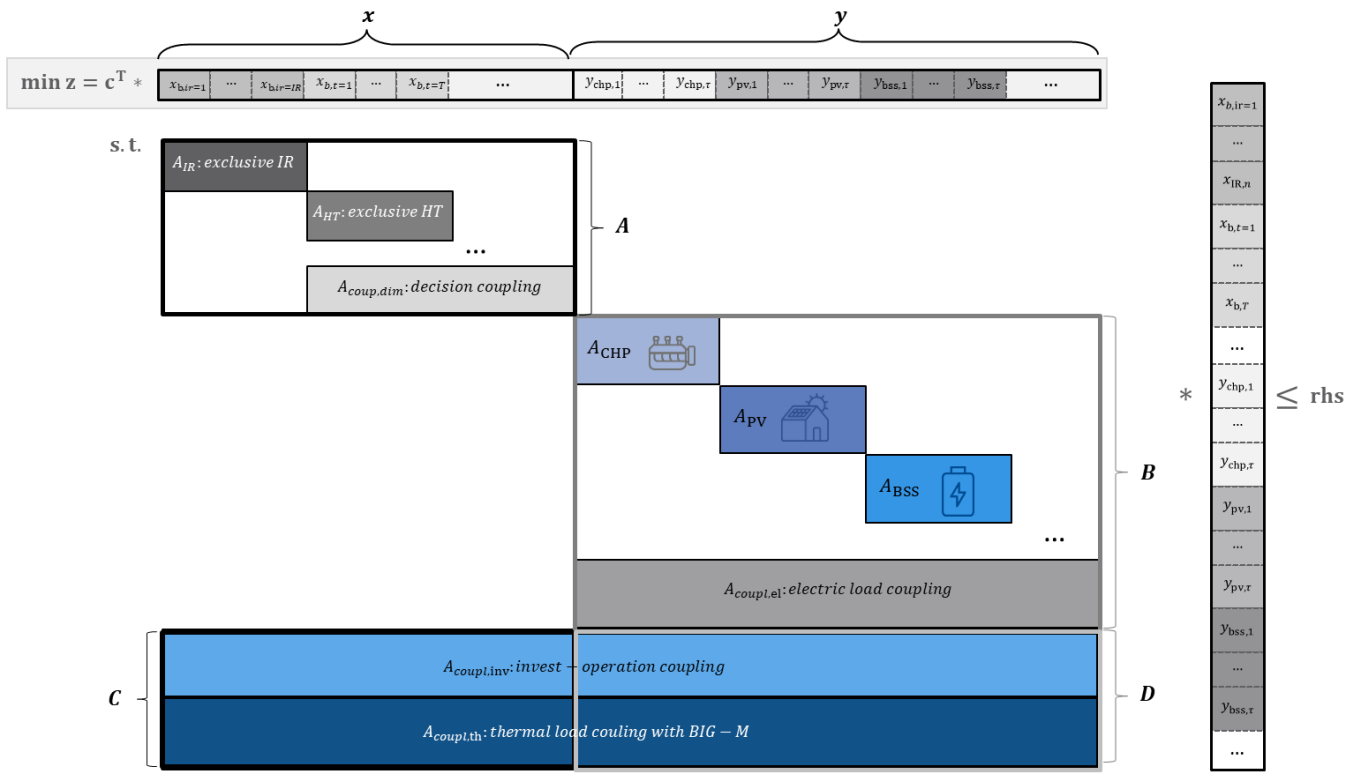

Figure 11. Blockangular structure of a building entity

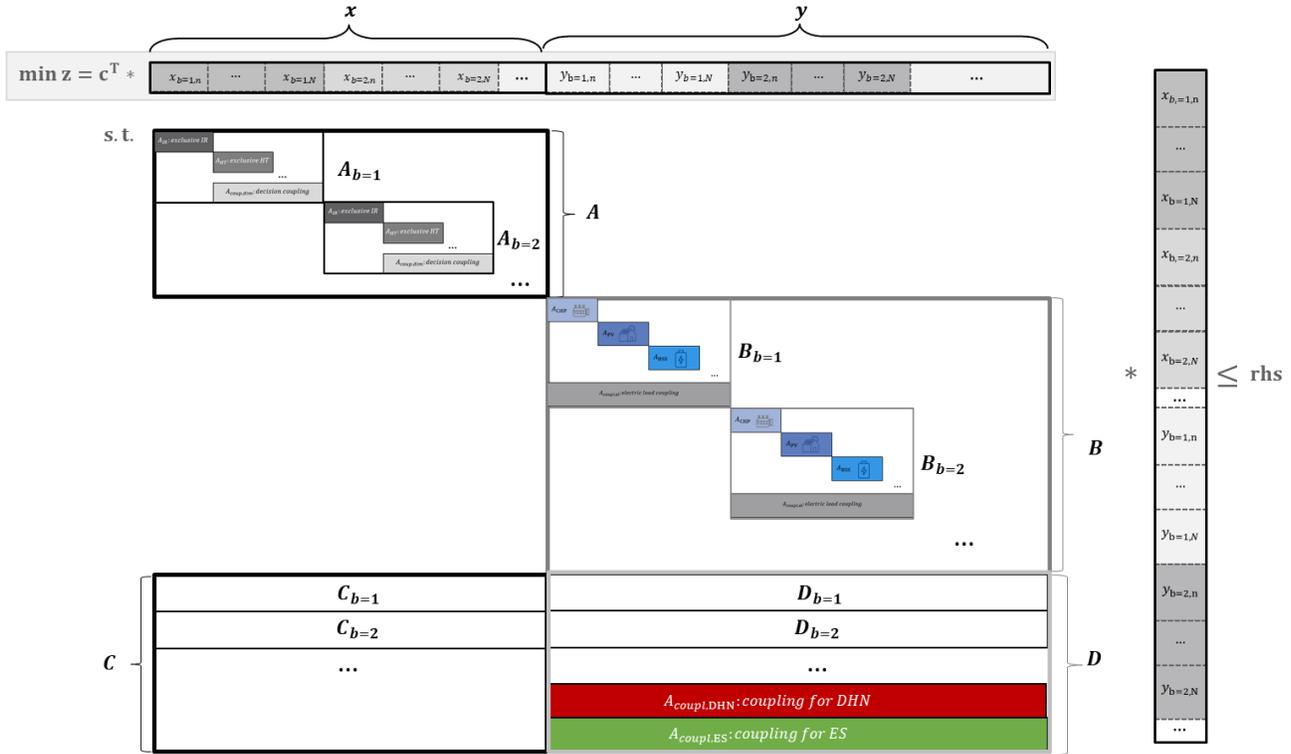

Figure 12. Blockangular structure of a group in the DMES connected via DHN and ES

In this formulation $x$ is describing the vector of independent variables of the first stage, the MP, that can be in the range of real and/or integer numbers. $y$ is describing the vector of independent variables of the second stage, the SP, in the range of real numbers only. The applied BD decomposes along the time stage (i.e. time of two depended decisions) [48].

$$\min z = c^T \cdot x + d^T \cdot y$$

*Formula 23.*

s.t.

$$A \cdot x \leq d_x$$

$$B \cdot y \leq d_y$$

$$C \cdot x + D \cdot y \leq a$$

$$x \in \mathbb{R} \vee \mathbb{Z}, y \in \mathbb{R}$$

In our case the first stage represents the decisions on installation and sizing of DHN pipelines and ECT, ESS and HE technologies, as well as the decision on IR and their respective restrictions, depicted in Formula 24.

$$\min z = c^T \cdot x$$

*Formula 24.*

s.t.

$$A \cdot x \leq d_x$$

$$x \in \mathbb{R} \vee \mathbb{Z}$$

The second stage represents the operation optimization including the decisions on the actual energy procurement and load coverage as well as the technology restrictions (see Formula 25).

$$\min w = d^T \cdot y$$

s.t.
$$B \cdot y \leq d_y$$

$$C \cdot x + D \cdot y \leq a$$

$$x \in \mathbb{R} \vee \mathbb{Z}, y \in \mathbb{R}$$

As the aim of the algorithm is to find the optimal solution of the complex problem, the two-stages must be linked. This linkage is created by introducing an additional independent variable Θ in the MP. Theta acts over the whole procedure as lower bound estimator for the objective value of the SP's in the MP, that changes its value in each iteration step. Therefore, the overall aim is to get the optimal objective of the MP. Besides this insertion, the SP's structure is merged with the separated coupling constraints of the input structure, with $W = [B; D]$, $T = [0; C]$, $h = [d_y; a]$. The alternated formulation of the problem follows Formulae 24 and 25:

Masterproblem:
$$\min z = c^T \cdot x + \Theta$$

s.t.
$$A \cdot x \leq d_x$$

$$x \in \mathbb{R} \vee \mathbb{Z}$$

Subproblem:
$$\min w = d^T \cdot y$$

s.t.
$$Wy + T\overline{x} \leq h$$

$$x \in \mathbb{R} \vee \mathbb{Z}, y \in \mathbb{R}$$

In the main phase of the procedure the MP and then the SP's are solved iteratively, as shown in the mid part of Figure 5. Each iteration contains one run of the first stage problem and a run of the second stage problem. The goal of the first stage problem is to decide upon the first stage decision variables, $x$ and Θ. The $x$ as representatives for investment decisions including the dimensioning of technologies, are then transferred to the second stage problem of the same iteration. This can be seen in beforementioned equation where $\overline{x}$ denotes the MP's optimal objective values in the actual iteration step. Θ is used as lower bound (LB) estimator for the real objective value of the SP's, as already introduced. The goal of the second stage problem is to find the expected value of the second stage problem as upper bound (UB) estimator, and the gradient of the first stage variable with respect to the second stage objective. LB and UB should converge over the whole iteration process (Note: they do not have to converge per step) if the optimization problem is convex [43, 44]. This reflects a better approximation of the operational optimization in every Benders iteration.

In the normal run, convergence of these two limits is checked every iteration as a stop criterion. If convergence is achieved the process is aborted and the optimal solution has been found. Otherwise cuts are introduced in the MP. These cuts, the so-called feasibility and optimality cuts, are restrictions that serve to communicate between MP and SP. Feasibility-cuts are generated if the calculation of the SP is not solvable. Thereby, the infeasibility of the SP is compensated by adding slack variables. This may be the case if the dimensioning of energy technologies found in the MP is not sufficient to serve the maximum heat and or electricity demands. This type of pruning is prevented in the present approach by the integration of slack variables into the technology operation models, which guarantee the feasibility of the problem. In our case these slack variables represent unlimited heat and electricity sources or sinks, with high penalty costs.

Therefore, only Optimality Cuts (OC) are used if a solution has been found but it does not satisfy the convergence criterion. OC successively restrict the solution space of the auxiliary variable θ in the master problem. The rationale of the restriction is based on the weak duality theorem and defines the SP objective solution as the lower bound for the value θ. The OC restriction is represented by the following Formula 28:

$$\theta \geq e_c - E_c * x$$

<div align="right">*Formula 28.*</div>

$$e_c = \sum_{w \epsilon(\Omega)} \pi * \lambda^{vT} * h$$

$$E_c = \sum_{w \epsilon(\Omega)} \pi * \lambda^{vT} * T$$

The variable $\lambda^{vT}$ denotes the dual solution of the problem $w\epsilon(\Omega)$. In Addition, we use Pareto-Optimal Cuts (POC) based on Magnanti and Wong [49]. This technique exploits the fact that there are several optimal dual solutions to the Benders subproblem. The different dual solutions have an effect on the quality of the OC. The strongest of all cuts is sought in accordance with Formula 29.

$$e_{Pc} - E_{Pc} * x \geq e_{OC} - E_{OC} * x$$

<div align="right">*Formula 29.*</div>

$$\max e_{Pc} - E_{Pc} * x$$

The POC chooses the dual solution that is closer to the core point of the MP solution space. A point is called a core point if it is in the relative interior of the solution space and the hull is convex [50]. This approach improves the convergence of the solution algorithm by efficiently narrowing down the space of investigation. The implementation of this technique follows the further development by Papadakos. In [50] it is proved that it is not necessary to know the core point of the solution space to generate a POC. For this approach, the solution space is iteratively sampled for the core point and thus further improved. The core point for the actual iteration $it$ is calculated from the core point of the previous iteration and the optimal solution of the MP, as in Formula 30.

$$CorePoint_{it} = \tau * CorePoint_{it-1} + (1 - \tau) * x_{opt,it}$$

<div align="right">*Formula 30.*</div>

The literature recommends choosing a control parameter $\tau$ of 0.5 (see e.g. [51]). Subsequently, an auxiliary subproblem of the BD is computed. This auxiliary subproblem has the same structure as the subproblem, but does not use the solution $x_{opt,t}$ of the MP for the calculation, but the core point $CorePoint_t$. The auxiliary subproblem generates a dual solution by means of which a POC similar to the OC can be determined for the MP. The combination of POC and OC application generates distinctive cuts that allow faster bounding of the solution space.

### Stage 3: Lagrange Relaxation

LR is a method suitable for the relaxation of problems with complicating constraint [52, 53]. This in the present case represent the operational couplings from Formulae 21 and 22 in the subproblems of the BD. LR is applied in the case that more than one building is connected via DHN and/ or ES. Therefore, we combine BD with the LR in a Cross-Decomposition approach. LR is applied to solve the Subproblem of a BD iteration. The process scheme of the applied Lagrange relaxation (see Figure 5, right part). is based on the concept from [54]. The procedure is divided into two convoluted loops, an outer and an inner loop. The inner loop representing the basic LR method, the outer being necessary to correct bounds of the model, if convergence of LR is not reached, due to the homogeneity of the operation optimization problem. The correction of the outer loop is based on the deviation of the relaxed constraint and represents a quantity adjustment. In the inner loop – the basic LR - a full separation of the operation problem in distinctive problems per building entity is achieved by relaxing the restrictive constraints and the integration of a Lagrange penalty term per relaxed coupling constraint using dual variables $\lambda \epsilon \mathbb{R}$ in the objective function [55]. This means that the LR dualizes the coupling constraint for heat and electricity exchange between the building entities. By this dualization the LR defines a master problem $L(y, \lambda_t^{th}, \lambda_t^{el})$ which is composed according to the Formula 31. The MP adjusted objective function includes the operating cost and the cost of LR, which is the result of the residuals $R_t^{el}, R_t^{th}$ multiplied by their shadow price $\lambda_t^{el}, \lambda_t^{th}$. The residuals are the sum of the deviation of the relaxed constraint.

$$L(y, \lambda_t^{th}, \lambda_t^{el}) = \min z = \sum_{b \in B} \sum_{j \in \{ECT,ESS,HE\}} \sum_{t \in T} \left( (1 - \varepsilon) \cdot \frac{y_{b,j,t} \cdot c_{b,j,t}^{OPEX}}{PVF_j} + \varepsilon \cdot y_{b,j,t} \cdot gwp_{j,t}^{Operation} \right)$$   *Formula 31.*

$$+ \sum_{t \in T} \left( \lambda_t^{el} \cdot \sum_{b \in B} (y_{b,EG,t}^{toES} - y_{b,EG,t}^{fromES}) \right)$$

$$+ \sum_{t \in T} \lambda_t^{th} \left( \sum_{b \in B} (y_{b,HE,t}^{toDHN} - y_{b,HE,t}^{fromDHN}) - \sum_{l \in L} \dot{Q}_{l,t}^{loss} \right)$$

s.t.
$$\sum_{b \in B} (y_{b,EG,t}^{toES} - y_{b,EG,t}^{fromES}) = R_t^{el}$$

$$\sum_{b \in B} (y_{b,HE,t}^{toDHN} - y_{b,HE,t}^{fromDHN}) - \sum_{l \in L} \dot{Q}_{l,t}^{loss} = R_t^{th}$$

The dual variables $\lambda$ are called the Lagrange multipliers and can be interpreted as a shadow price of energy purchase/ sale. In this MP the decision variables $y_{b,EG,t}^{toES}$, $y_{b,EG,t}^{fromES}$, $y_{b,HE,t}^{toDHN}$, and $y_{b,HE,t}^{fromDHN}$ represent quantity bids from the respective operation optimization results of the building entities. In a first iteration these are based on a high marginal price for energy procurement/ energy sales, on the basis of the levelized costs for energy (LCOE) of a single building optimization. This is the result of a first rudimentary optimization within the preprocessing. Based on the Lambdas the objective functions of each separate operational problem are updated, which reflects a price signal for energy sharing. After the subproblems have been optimized, the residuals of the violated relaxed constraint are determined for each time step. The residuals are used to evaluate the convergence of the LR iterative process based on the upper bound $UB_{it}$, lower bound $LB_{it}$ and their difference $Gap_{it}$ according to Formulae 32 – 34 with $\lambda_{max}^{el}$, $\lambda_{max}^{th}$ the maximum shadow prices in all LR iterations.

$$UB_{it} = \min(UB_{it-1}, \sum_{t \in T} (R_t^{el} \cdot \lambda_{max}^{el} + R_t^{th} \cdot \lambda_{max}^{th}))$$   *Formula 32.*

$$LB_{it} = \max(LB_{it-1}, z)$$   *Formula 33.*

$$Gap_{it} = \frac{UB_{it} - LB_{it}}{UB_{it}}$$   *Formula 34.*

The difference $Gap_{it}$ is subsequently taken to see if it is smaller than a predefined gap $\epsilon_{Gap}$ (see Formula 35). Convergence is reached if this applies true or if there has been no improvement within the last iterations (see Formula 36) and all-time steps have been sufficiently investigated, thus having a residual below a threshold $\epsilon_{Residual}$ (see Formula 37).

$$Gap_{it} < \epsilon_{Gap}$$   *Formula 35.*

$$Gap_{it-3} = Gap_{it}$$   *Formula 36.*

$$R_t > \epsilon_{Residual}$$   *Formula 37.*

If the inner loop converges, the partial results of the subproblem are assembled and the optimal objective function value, the dual solutions and the residual are passed to the outer loop. Convergence of the outer loop is only reached if only Formula 35 applies.

If the inner loop does not converge, the values for $\lambda_t^{el}$, $\lambda_t^{th}$ are updated. The lambdas are only adjusted if the residual in this time step $t$ is greater than the tolerance as already introduced in Formula 37. Adaption is the task of changing the lambdas in such a way that equilibrium is reached in the relaxed constraint. This relates to a price adoption. For this purpose, a sub-gradient method, the resilient backpropagation algorithm (Rprop) is used [56] . The Rprop uses a step-size procedure in the search for the optimal coordination prices $\lambda_t$ as shown in Formula 38,

$$\lambda_t = \lambda_t + \Delta_{t,it}$$

<div align="right">*Formula 38.*</div>

$$\Delta_{t,it} = \nu_{t,it} \cdot \alpha, \quad if \ it = 1$$

<div align="right">*Formula 39.*</div>

$$\Delta_{t,it} = \nu_{t,it} * \begin{cases} \eta^+ \cdot \Delta_{t,it-1}, & \nu_{t,it-1} * \nu_{t,it} > 0 \\ \eta^- \cdot \Delta_{t,it-1}, & \nu_{t,it-1} * \nu_{t,it} < 0, \quad if \ it > 1 \\ 0, & else \end{cases}$$

<div align="right">*Formula 40.*</div>

In the first iteration step it = 1, the price adjustment $\Delta_{t,it}$ is determined using the sign $\nu_{t,it}$ of the residual according to formula 39. For the further iterations, a case discrimination as shown in formula 40 is done. In the case that the sign of the residual of the current and the previous iteration are the same, a price change with the acceleration factor $\eta^+$ is carried out. If the signs differ, the price was adjusted beyond the optimal adjustment. The step size is reduced with the braking factor $\eta^-$. In the case that a residuum equals zero, the price is not adjusted. The literature recommends the following parameterisation for the initialisation step size $\alpha = 0.1$, the acceleration factor $\eta^+ = 1.0$ and the braking factor $\eta^-=1.2$ [56]. In addition, to the Rprop we use a shifting procedure if the lambdas exceed a critical value, i.e. the maximum generation costs per technology. If this applies $\lambda_t$ is set to the costs of the slack variables in the next iteration. This would also happen without a shift procedure, but would require more iterations. Therefore, the shift procedure accelerates the convergence of the inner loop.

*Parallelization*

For the application of the decomposition approaches a splitting of the partial and sub-problems on distributed computers is advantageous. The approach chosen in this work is particularly well suited for parallelization, since multiple decompositions of the problem take place: parallelization can start at different levels, as shown in Figure 13. In Ga, the fitness calculation can be calculated as per chromosome. A chromosome is subdivided into groups that can in turn be processed in parallel. Depending on the size of the groups, arbitrarily complex MILPs result. In the example shown, two groups are formed out of the top chromosome, and five out of the bottom chromosome. The latter are solved as single building optimizations, which are less complex in comparison to the problems defined by groups with several buildings and a coupling through DHN and ES. The interleaved master, slave architecture in our parallelization approach allows problems to be dynamically assigned to distributed resources. Thus, the more complex problems are prioritized when computational resources are limited. Both individual building optimization and group optimization can be further decomposed by the varying weights of the applied eta-weight method at the BD level. The problems of the groups with the restrictive constraint can furthermore be split by LR and solved in separate instances. This results in a total of three levels of multi-level parallelization.

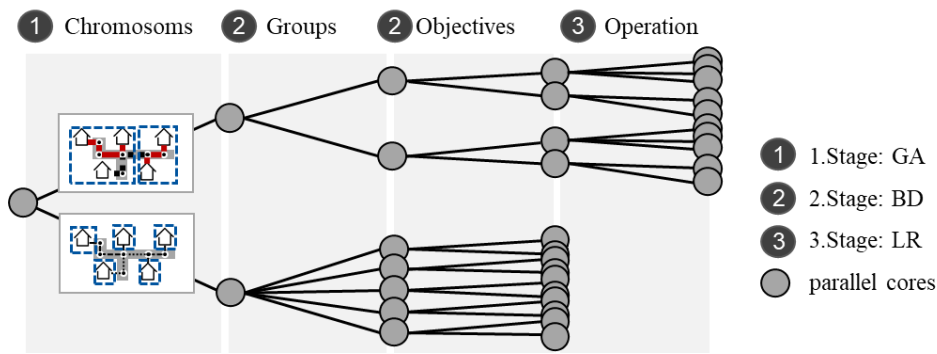

Figure 13. Scheme of the multi-level parallelization

## 4. SUMMARY AND FUTURE WORK

In this paper, a method for the stakeholder-oriented optimization of DMES is presented. This method is characterized by a high detail depth related to contextual level, including a house sharp spatial resolution, a broad technical coverage, as well as market and services coverage, e.g. self-consumption and energy sharing. This level of detail drives the complexity of the program and thus the use of computational resources. To overcome this problem, we present a mathematical formulation that splits the problem in the time and system domains. This decomposition using a GA, BD and LR moreover allows a multi-level parallelization of the problem in order to accelerate its computation. Overall, the presented approach allows to gain important techno-economic and -ecologic insights in to future design options for DMES in the scale of city districts, with around 50 to 80 buildings. The depth within the method facilitates the practical transfer of the identified measures. The full paper of this preprint aims to show the promising advantages of the method on the basis of exemplary results.

Further research work is planned, in particular, to improve the LR. Based on the literature, there are further possibilities to accelerate the computation time by using a hybrid combination of LR and Dantzig Wolfe decomposition. In theory this hybrid combination aims to inherit the information of all local energy market bids in the cyclic LR in the single building operational optimization. This approach may allow for a faster convergence of the LR, so that the outer loop of the current approach may be obsolete.

The presented method is currently based on an operational representation covering one typical year. The size of this period is especially important for the fact that the design-relevant times (e.g. peak room-heating demand in winter) for technologies and time-dependent/seasonal behaviour by varying external influences such as temperature and solar radiation are considered in the DMES optimization. Methods for time-series aggregation (incl. time-series clustering) promise a possibility to narrow down the period to a few typical days, which represent the year without a large loss of information. As with the decomposition methods, this promises additional depletion of computer resource usage. However, time-series aggregation and the decomposition approaches are not necessarily compatible. This will also be investigated in the future.

## 5.  FUNDING REFERENCE

This work was developed in part within the framework of the project "EnEff:Stadt:Q-SWOP" funded by the Federal Ministry for Economic Affairs and Climate Action of Germany under grant number 03ET1587B.